\documentclass[aps,twocolumn,superscriptaddress,showpacs,showkeys]{revtex4}
\usepackage{graphics,graphicx,dcolumn,bm,fleqn,epic,eepic,float}
\usepackage{amssymb,amsmath,multirow,rotate,color,float,times}
\usepackage{color}
\usepackage{soul}                    
\definecolor{red}{rgb}{1,0,0}
\definecolor{green}{rgb}{0,1,0}
\definecolor{blue}{rgb}{0,0,1}

\newcommand{\bite}{\begin{itemize}}
\newcommand{\eite}{\end{itemize}}
\newcommand{\benu}{\begin{enumerate}}
\newcommand{\eenu}{\end{enumerate}}
\newcommand{\bverb}{\begin{verbatim}}
\newcommand{\everb}{\end{verbatim}}

\newcommand{\beq}{\begin{equation}}
\newcommand{\eeq}{\end{equation}}
\newcommand{\barr}{\begin{array}}
\newcommand{\earr}{\end{array}}
\newcommand{\p}{\partial}
\newcommand{\pr}{\prime}

\newcommand{\Four}{\mathbf{F}}
\newcommand{\DDrift}{D^{(1)}}
\newcommand{\DDiff}{D^{(2)}}
\newcommand{\xosig}{\frac{x}{\sigma}}
\newcommand{\sigom}{\sigma\omega}
\begin{document}

\title{Analysis of stochastic time series in the presence of strong measurement noise}

\author{B.~Lehle}
\affiliation{Hohewartstr.~44, D-70469 Stuttgart, Germany}

\begin{abstract}
A new approach for the analysis of Langevin-type stochastic processes in the presence of strong measurement noise is
presented. For the case of Gaussian distributed, exponentially correlated measurement noise it is possible to
extract the strength and the correlation time of the noise as well as polynomial approximations of the
drift and diffusion functions from the underlying Langevin equation.
\end{abstract}

\pacs{05.40.Ca,  
      02.50.Ey}  

\keywords{Measurement noise, Stochastic processes}

\maketitle

\section{Introduction}
\label{sec:intro}

In the last years there has been significant progress in the analysis and characterization
of the dynamics of processes underlying the time series of
complex dynamical systems \cite{schreiberbook,abarbanel,friedrich08}.

If the temporal evolution of a quantity $X_t$ can be described by a Langevin equation, it is possible to extract drift
and diffusion functions of the underlying stochastic process from a given time series.
This can be done because the moments of the conditional probability densities of $X_{t+\tau}|_{X_t=x}$
can be related to these functions.

Since this approach was introduced \cite{friedrich97,ryskin97,siegert98,friedrich00,gradisek00,friedrich08}
it has been successfully carried out in a broad range of fields. For example for data from financial markets
\cite{finance00}, traffic flow \cite{trafic02}, chaotic electrical circuits \cite{circuit03,circuit04}, human heart beat
\cite{heart04}, climate indices \cite{climate04,climate05}, turbulent fluid dynamics \cite{fluid07}, and for
electroencephalographic data from epilepsy patients \cite{epilepsy08,epilepsy09}.

Real-world data, however, also give rise to some problems. One of them is, that experimental data are only given with a
finite sampling rate. So methods had to be proposed to deal with the effects arising from this
fact \cite{sampling02,sampling05,sampling08}.

Another problem is the virtually unavoidable measurement noise \cite{schreiberbook,noise93,noise00,sampling08}.
In the presence of measurement noise $Y_t$ the values of $X_t$ or any of its probability densities are
no longer accessible, but only $X^*_t=X_t+Y_t$ and {\em its} density distributions.

Recently an approach has been presented which allows the estimation of drift and diffusion functions
in the presence of strong delta-correlated, Gaussian noise \cite{iterative06,iterative10}.
Starting with initial estimates for the noise strength and the drift and diffusion functions a functional
of these unknowns is iteratively minimized.

The aim of this paper is to introduce an alternative approach to the treatment of strong measurement noise. This
approach is able to deal also with exponentially correlated, Gaussian noise.
The basic idea is, not to look at the conditional
moments in the first place but at the joint probability density $\rho(x_1,x_2,\tau)$
of pairs $(X_t,X_{t+\tau})$.
If the measurement noise is independent of $X_t$, then $(X_t,X_{t+\tau})$ and
$(Y_t,Y_{t+\tau})$ are independent random variables and the joint probability density
$\rho^*(x_1,x_2,\tau)$ of their sum $(X^*_t,X^*_{t+\tau})$ is given by the convolution of $\rho$ and $\rho_Y$, where
$\rho_Y(x_1,x_2,\tau)$ is the joint probability density of $(Y_t,Y_{t+\tau})$.

The noise is assumed to be Gaussian and the Gauss function has some special properties with regard to
convolution and Fourier transform. This makes it possible to
extract the noise parameters from the moments of $\rho^*$. Furthermore the abovementioned relation
between the conditional moments and the unknown functions can be transformed into a relation in Fourier space.
This allows polynomial approximations of the drift and diffusion functions to be extracted using
purely algebraic relations between quantities that can be calculated directly from a given, noisy time series.

This paper is organized as follows: Section \ref{sec:process} is devoted to the noise-free stochastic process, the
definition of its joint probability density and expressions for the moments of this density in terms of a Taylor-It\^o
expansion. Section \ref{sec:noise} provides the properties of the measurement noise under consideration and in section
\ref{sec:noisy_process} expressions for the moments of a noisy process will be derived. In section \ref{sec:extracting noise}
these expressions will be used to extract the parameters of the measurement noise and in
section \ref{sec:extracting coeffs} to extract polynomial approximations for drift and diffusion functions.
Finally in section \ref{sec:examples} the results are applied to some synthetic time series. The used properties
of the Gauss function and further computational details are given in appendices~\ref{app:gauss} to
\ref{app:err_sigma}.

\section{Stochastic process}
\label{sec:process}

Let $X_t$ be a stochastic process that can be described by a time-independent It\^o Langevin equation
\begin{equation}
dX_t = \DDrift(X_t)dt+\sqrt{\DDiff(X_t)}dW_t \label{langevin},
\end{equation}
where $\DDrift$ and $\DDiff$ are the Kramers-Moyal coefficients of the corresponding Fokker-Planck equation and $dW$ is
the increment of a standard Wiener process ($<dWdW>=dt$).

The above equation cannot be solved analytically in general. To get finite-time step approximations for the temporal
evolution of $X_t$, a Taylor-It\^o expansion can be applied.
\begin{equation}
X_{t+\tau}=X_t+h(X_t,\tau)+R\label{taylor_ito}
\end{equation}
The function $h$ contains the expansion terms and $R$ is a remainder that will be dropped when using the approximation.
The above equation allows the numerical integration of Eq.~(\ref{langevin}) but will not be needed in the following. What will
be needed are approximations for the first and second moment of $X_{t+\tau}-X_t$.
\begin{subequations}
\begin{eqnarray}
<(X_{t+\tau}-X_t)|_{X_t=x}> &=& h_1(x,\tau) +R_1 \\
<(X_{t+\tau}-X_t)^2|_{X_t=x}> &=& h_2(x,\tau) +R_2
\end{eqnarray}\label{expansion_moments}\end{subequations}
A Taylor-It\^o expansion also provides expressions for these moments. For an expansion of order $k$ the functions $h_1$ and
$h_2$ are polynomials of order $k$ in $\tau$ (without a constant part) and the remainders $R_1$ and $R_2$ are of
order $\tau^{k+1}$.
\begin{subequations}
\begin{eqnarray}
h_1(x,\tau) &=& c_1(x)\tau+\ldots+c_k(x)\tau^k \label{expansion_moments_poly_h1}\\
h_2(x,\tau) &=& d_1(x)\tau+\ldots+d_k(x)\tau^k
\end{eqnarray}\label{expansion_moments_poly}\end{subequations}
The polynomial coefficients generally are functions of $\DDrift$ and $\DDiff$ and their derivatives. The first order
approximations that will be used later are given by
\begin{subequations}
\begin{eqnarray}
h_1(x,\tau) &=& \DDrift(x)\tau \\
h_2(x,\tau) &=& \DDiff(x)\tau.
\end{eqnarray}\label{expansion_moments_euler}\end{subequations}
A detailed description of the Taylor-It\^o expansion and its moments can be found in \cite{platen99}.

The probability density of $X_t$ and the joint probability density of $(X_t,X_{t+\tau})$ are formally given by
\begin{subequations}
\begin{eqnarray}
\rho(x_1) &=& \lim_{T\to\infty}\frac{1}{T}\int_0^T\delta(X_t-x_1)dt\\
\cr
\rho(x_1,x_2,\tau) &=& \lim_{T\to\infty}\frac{1}{T}\int_0^T\delta(X_t-x_1)\cr
\cr
                  && \qquad\quad \times\quad \delta(X_{t+\tau}-x_2)dt.\label{dens2_def}
\end{eqnarray}\end{subequations}
Let the conditioned moments $m_j$ be defined as
\begin{eqnarray}
m_j(x_1,\tau) &=& \int_{-\infty}^{\infty}(x_2-x_1)^j\rho(x_1,x_2,\tau)dx_2.
\end{eqnarray}
(These moments are not to be confused with the conditional moments which usually are defined as $m_j/m_0$).
Inserting Eq.~(\ref{dens2_def}) the moments can be expressed as
\begin{eqnarray}
m_j(x_1,\tau) &=& \lim_{T\to\infty}\frac{1}{T}\int_0^T\delta(X_t-x_1)\cr
\cr
                  && \qquad\quad \times\quad (X_{t+\tau}-x_1)^jdt. \label{m_j_no_noise}
\end{eqnarray}
Using
\begin{eqnarray}
\lim_{T\to\infty}\frac{1}{T}\int_0^T\delta(X_t-x_1)f(X_t)dt =\qquad\qquad&&\cr
\quad<f(X_t)|_{X_t=x_1}>\rho(x_1)&& 
\end{eqnarray}
leads to
\begin{eqnarray}
m_j(x_1,\tau) &=& <(X_{t+\tau}-X_t)^j|_{X_t=x_1}>\rho(x_1)
\end{eqnarray}
Inserting Eqs.~(\ref{expansion_moments}) and omitting the remainders the final approximations of the moments up to order two
are therefore given by
\begin{subequations}
\begin{eqnarray}
m_0(x_1) &=& \rho(x_1)\\
\cr
m_1(x_1,\tau) &=& h_1(x_1,\tau) \hspace{2pt} m_0(x_1)\\
\cr
m_2(x_1,\tau) &=& h_2(x_1,\tau) \hspace{2pt} m_0(x_1)
\end{eqnarray}\label{constituting}\end{subequations}
These are the equations that establish the relation between the observable quantities $m_j$ and the unknown coefficients
contained in $h_1$ and $h_2$ within the precision of the chosen Taylor-It\^o expansion.

In the absence of noise Eqs.~(\ref{constituting}) can directly be used to determine $\DDrift$ and $\DDiff$. If e.g. the
above mentioned first order scheme was chosen, the equations read
\begin{eqnarray*}
\frac{m_1(x_1,\tau)}{ m_0(x_1)} &=& \tau\DDrift(x_1)+O(\tau^2)\\
\cr
\frac{m_2(x_1,\tau)}{ m_0(x_1)} &=& \tau\DDiff(x_1)+O(\tau^2).
\end{eqnarray*}

\section{Measurement noise}
\label{sec:noise}

Let $Y_t$ be exponentially correlated Gaussian noise as produced by an Ornstein-Uhlenbeck process.
\begin{eqnarray}
dY_t &=& -aY_t dt+b\hspace{2pt}dW_t
\end{eqnarray}
This equation can be solved analytically (see e.g. \cite{risken89}). The probability densities of $Y_t$ can be expressed
either in terms of $a$ and $b$ or, equivalently, in terms of the `macroscopic' parameters $T$ (characteristic
time scale) and $\sigma^2$ (variance) with
\begin{eqnarray}
T &=& \frac{1}{a}\quad\hbox{and}\quad \sigma^2 = \frac{b^2}{2a}.
\end{eqnarray}
The unconditioned distribution of $Y_t$ will be denoted by $K$ because it will serve as a convolution kernel later.
It is given by
\begin{eqnarray}
K(x) &=& \frac{1}{\sqrt{2\pi\sigma^2}}e^{-\frac{1}{2} \left( \frac{x}{\sigma} \right)^2 }.
\end{eqnarray}
The joint probability density of $(Y_t,Y_{t+\tau})$ can then be expressed as
\begin{eqnarray}
\rho_Y(x_1,x_2,\tau) &=& K(x_1) \frac{1}{\sqrt{2\pi s^2}}
                          e^{-\frac{1}{2}\left(\frac{x_2-\mu(\tau)x_1}{s}\right)^2},\label{pdf_noise}
\end{eqnarray}
where the decay-function
\begin{eqnarray}
\mu(\tau) &=& e^{-\frac{\tau}{T}}
\end{eqnarray}
and the auxiliary quantity
\begin{eqnarray}
s^2 &=& \sigma^2(1-\mu^2(\tau))\label{s_sigma}
\end{eqnarray}
have been introduced for notational simplicity. The parameter $T$ will also be called `the' correlation time.
This is motivated by
\begin{eqnarray}
\frac{<Y_t Y_{t+\tau}>}{\sigma^2} &=& e^{-\frac{\tau}{T}}.
\end{eqnarray}

\section{Noisy stochastic process}
\label{sec:noisy_process}

Let $X^*_t=X_t+Y_t$ be the sum of a stochastic signal $X_t$ and measurement noise $Y_t$ as introduced in
sections~\ref{sec:process} and \ref{sec:noise}, respectively. Then, because $X$ and $Y$ are independent,
the joint probability density $\rho^*$ of the pairs of noisy variables, $(X^*_t,X^*_{t+\tau})$, is given by the
convolution of $\rho$ and $\rho_Y$.
\begin{eqnarray}
\rho^*(x_1,x_2,\tau) &=& \rho_Y(x_1,x_2,\tau)*\rho(x_1,x_2,\tau)\cr
\cr
                     &=& \int_{-\infty}^\infty\int_{-\infty}^\infty\rho_Y(x_1-x^\pr_1,x_2-x^\pr_2,\tau)\cr
\cr
&&\times\quad \rho(x^\pr_1,x^\pr_2,\tau)dx^\pr_1dx^\pr_2 \label{pdf_noisy}
\end{eqnarray}
Instead of the conditioned moments $m_j$ only their noisy counterparts $m_j^*$ can be determined.
\begin{eqnarray}
m_j^*(x_1,\tau) &=& \int_{-\infty}^{\infty}(x_2-x_1)^j\rho^*(x_1,x_2,\tau)dx_2\label{m_j_eq_1}
\end{eqnarray}
These can be expressed as follows (see appendix~\ref{app:cond_mom_1}). The index of the variable $x_1$ and the function
argument of $\mu(\tau)$ are omitted for notational simplicity.
\begin{subequations}
\begin{eqnarray}
m^*_0(x)      &=& K(x)*m_0(x)\\
\cr
m^*_1(x,\tau) &=& K(x)*m_1(x,\tau)\cr
\cr
              && -(1-\mu)(xK(x))*m_0(x)\\
\cr
m^*_2(x,\tau) &=& K(x)*m_2(x,\tau)\cr
\cr
              &&-2(1-\mu)(xK(x))*m_1(x,\tau)\cr
\cr
              &&+(1-\mu)^2(x^2K(x))*m_0(x)\cr
\cr
              &&+s^2K(x)*m_0(x).
\end{eqnarray}\end{subequations}
These relations have already be derived earlier for the case of delta-correlated noise and have been formulated with
conditional probabilities \cite{iterative06}. The above convolutional notation makes it more obvious how to
proceed further. Because $K$ is a Gauss function with variance $\sigma^2$, the terms $x^jK(x)$ can be expressed as
(see appendix~\ref{app:gauss}):
\begin{eqnarray*}
xK(x) &=& -\sigma^2\p_xK(x)\\
x^2K(x) &=& (\sigma^2+\sigma^4\p^2_x)K(x)
\end{eqnarray*}
Because of $(\p_xf)*g=\p_x(f*g)$ this leads to (using Eq.~(\ref{s_sigma}) and omitting arguments
now completely):
\begin{subequations}
\begin{eqnarray}
m^*_0      &=& K*m_0 \label{eq28a}\\
\cr
m^*_1 &=& K*m_1 +(1-\mu)\sigma^2\p_x(K*m_0) \label{eq28b}\\
\cr
m^*_2 &=& K*m_2 +2(1-\mu)\sigma^2\p_x(K*m_1)\cr
\cr
              &&+(1-\mu)^2\sigma^4\p^2_x(K*m_0)\cr
\cr
              &&+2(1-\mu)\sigma^2K*m_0
\end{eqnarray}\end{subequations}
Substituting $K*m_0$ and $K*m_1$ and using Eqs.~(\ref{eq28a}) and (\ref{eq28b}) yields
\begin{subequations}
\begin{eqnarray}
m^*_0      &=& K*m_0\\
\cr
m^*_1 &=& K*m_1 +(1-\mu)\sigma^2\p_xm^*_0\\
\cr
m^*_2 &=& K*m_2 +2(1-\mu)\sigma^2\p_xm^*_1\cr
\cr
              &&-(1-\mu)^2\sigma^4\p^2_xm^*_0 +2(1-\mu)\sigma^2m^*_0.
\end{eqnarray}\end{subequations}
Using Eqs.~(\ref{constituting}) to express $m_1$ and $m_2$ finally gives
\begin{subequations}
\begin{eqnarray}
m^*_0      &=& K*m_0\\
\cr
m^*_1 &=& K*(h_1m_0) +(1-\mu)\sigma^2\p_xm^*_0 \label{noisy_m1}\\
\cr
m^*_2 &=& K*(h_2m_0) +2(1-\mu)\sigma^2\p_xm^*_1\cr
\cr
              &&-(1-\mu)^2\sigma^4\p^2_xm^*_0 +2(1-\mu)\sigma^2m^*_0.
\end{eqnarray} \label{noisy_moments}\end{subequations}
Eq.~(\ref{noisy_m1}) can be used to extract the parameters of the measurement noise without determining the
Kramers-Moyal coefficients $\DDrift$ and $\DDiff$. This will be done in the next section.

\section{Extracting measurement noise parameters}
\label{sec:extracting noise}

Multiplying Eq.~(\ref{noisy_m1}) by a weight function $\Psi(x)$ and subsequently performing an integration with
respect to $x$ leads to
\begin{eqnarray}
A(\tau) &=& \int_{-\infty}^\infty(K(x)*(h_1(x,\tau)m_0(x))\Psi(x) dx\cr
        &&\qquad + (1-\mu(\tau))\sigma^2 B \label{extract_noise_1}
\end{eqnarray}
with
\begin{eqnarray}
A(\tau) &=& \int_{-\infty}^\infty m^*_1(x,\tau)\Psi(x)dx
\end{eqnarray}
and
\begin{eqnarray}
B &=& \phantom{-}\int_{-\infty}^\infty (\p_x m^*_0(x))\Psi(x)dx\cr
\cr
  &=& -\int_{-\infty}^\infty m^*_0(x)(\p_x \Psi(x))dx
\end{eqnarray}
In the last step an integration by parts has been applied, assuming $\Psi m^*_0 \to 0$ when $|x|\to\infty$.
Using Eq.~(\ref{expansion_moments_poly_h1}) the remaining integral
in Eq.~(\ref{extract_noise_1}) can be expressed as a polynomial in $\tau$. 
\begin{eqnarray}
&&\int_{-\infty}^\infty\{K(x)*(h_1(x,\tau)m_0(x)\}\Psi(x) dx\cr
&&\quad   = \sum_{j=1}^k\tau^j \int_{-\infty}^\infty\{K(x)*(c_j(x)m_0(x))\}\Psi(x) dx\cr
&&\quad   = \sum_{j=1}^k\tau^j \tilde C_j
\end{eqnarray}
Finally, after a division by $B$, Eq.~(\ref{extract_noise_1}) reads
\begin{eqnarray}
z(\tau) &=& (1-\mu(\tau))\sigma^2+C_1\tau+C_2\tau^2+\ldots \label{extract_noise_2}
\end{eqnarray}
with
\begin{eqnarray}
z(\tau) &=& \frac{A(\tau)}{B}\quad\hbox{and}\quad C_j\;=\;\frac{\tilde C_j}{B}.
\end{eqnarray}
Once a weight function $\Psi$ has been chosen, $z(\tau)$ can be derived from experimental data.
Eq.~(\ref{extract_noise_2}) then can be used to fit the unknown parameters $\sigma^2$, $T$ (contained
in $\mu=e^{-\frac{\tau}{T}}$) and $C_j$ on the right-hand side. This can be done for example by an
iterated least-square fit.

The simplest choice $\Psi=x$ is not the most accurate one, especially in the presence of heavy tails. It puts a
high weight on the tails of $m_1^*$, where
density is low and fluctuations of experimental data thus high. Choosing $\p_x\Psi$ to be a rough, piecewise
steady approximation of the density $m_0^*$ overcomes this problem and also assures $B\ne0$.

\begin{figure}[ht]
\begin{center}
\includegraphics*[width=8.6cm,angle=0]{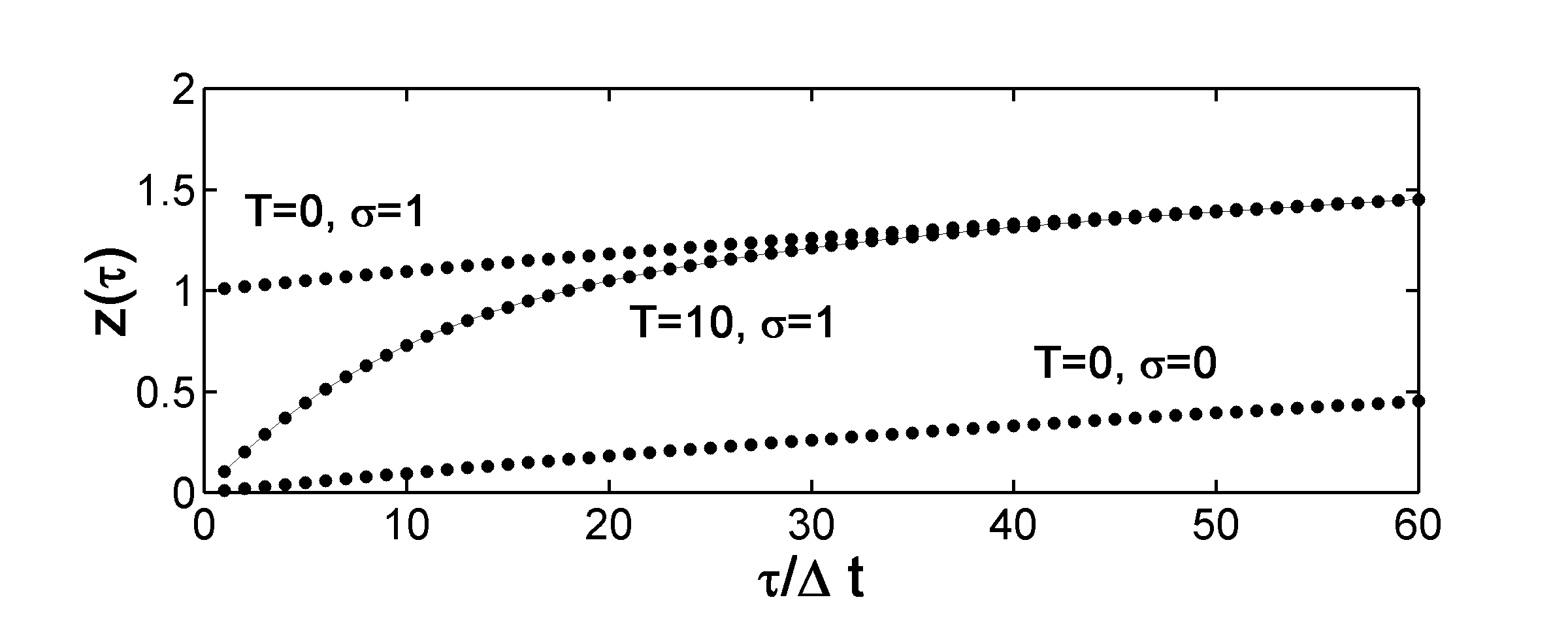}
\end{center}
\caption{\protect Values $\tilde z$ derived from experimental data (Ornstein-Uhlenbeck process) with and without
measurement noise ($\sigma=1$ and $\sigma=0$ respectively).
The correlation time $T$ is given in units of $\Delta t$. The process time scale is $100 \Delta t$. 
}
\label{fig0}
\end{figure}

No binning has to be applied in the evaluation of $z(\tau)$, if the values of $X^*$ are given at constant time-increments
$\Delta t$. Let $x_i=X^*(t_i)$ with $t_i=t_0+i\Delta t$. Then, instead of using box-counting techniques to
approximate the density $\rho^*$, the temporal evolution of $X^*_t$ itself can be approximated by a piecewise
constant function:
\begin{eqnarray}
\tilde X^*_t &=& x_i \qquad (t_i\le t <t_{i+1})
\end{eqnarray}
The density distribution of $(\tilde X^*_t,\tilde X^*_{t+\tau_k})$ with $\tau_k=k\Delta t$ can be evaluated
using Eq.~(\ref{dens2_def}) and
is given by a sum of Dirac-distributions (the coordinates $x_1^\pr$ and $x_2^\pr$ are used to avoid confusion
with the values $x_i$ of the given time series).
\begin{eqnarray}
\tilde \rho^*(x_1^\pr,x_2^\pr,\tau_k) &=& \frac{1}{N}\sum_{i=1}^N \delta(x_i-x_1^\pr)\delta(x_{i+k}-x_2^\pr)\label{approx_dens}
\end{eqnarray}
The moments of $\tilde\rho^*$ evaluate to
\begin{eqnarray}
\tilde m_j^*(x_1^\pr,\tau_k) &=& \frac{1}{N}\sum_{i=1}^N (x_{i+k}-x_1^\pr)^j\delta(x_i-x_1^\pr)\label{approx_mom}
\end{eqnarray}
what results in
\begin{eqnarray}
\tilde A(\tau_k) &=& \frac{1}{N}\sum_{i=1}^N (x_{i+k}-x_i)\Psi(x_i)\cr
\tilde B &=& -\frac{1}{N}\sum_{i=1}^N \Psi^\pr(x_i)
\end{eqnarray}
where $\Psi^\pr$ denotes the derivative with respect to $x$. The values $\tilde z(\tau_k)$ derived from a given
time series $x_i$ can therefore be expressed as
\begin{eqnarray}
\tilde z(\tau_k) &=& -\frac{\sum_i (x_{i+k}-x_i)\Psi(x_i)}{\sum_i \Psi^\pr(x_i)}.\label{approx_z}
\end{eqnarray}
To be able to successfully perform a fit it must be possible to distinguish
between the polynomial $\sum C_j\tau^j$ and the function $1-e^{-\frac{\tau}{T}}$ within the given range
of increments $\tau$. This means that a good polynomial approximation for $e^{-\frac{\tau}{T}}$ should require
a higher order than the polynomial defined by the $C_j$.
The proposed method is therefore limited to measurement noise with a
correlation time $T$ considerably smaller than the time scale of the underlying stochastic process.
An illustrative example is given in Fig.~\ref{fig0}.

\section{Extracting drift- and diffusion coefficients}
\label{sec:extracting coeffs}

Going back to Eqs.~(\ref{noisy_moments}) and putting the convolution terms on the right-hand side gives
\begin{subequations}
\begin{eqnarray}
&&m^*_1 - (1-\mu)\sigma^2\p_xm^*_0 = K*(h_1m_0)\\
\cr
&&m^*_2 - 2(1-\mu)\sigma^2(\p_xm^*_1+m^*_0)\cr
\cr
&&\qquad +(1-\mu)^2\sigma^4\p^2_xm^*_0  = K*(h_2m_0).\phantom{\int}
\end{eqnarray}\label{space_m1_m2}\end{subequations}
To get rid of the derivatives of $m^*_j$ a Fourier transform
\begin{eqnarray}
\Four:\quad f(x)\to \hat f(\omega)=\int_{-\infty}^\infty e^{-i\omega x}f(x)dx
\end{eqnarray}
is applied. In Fourier space Eqs.~(\ref{space_m1_m2}) now read
\begin{subequations}
\begin{eqnarray}
\hat m^*_1 - M i\omega\hat m^*_0 \qquad\qquad&=& \gamma_1 \label{Four_m1}\\
\hat m^*_2 - 2M(\hat m^*_0 +i\omega\hat m^*_1)-M^2\omega^2\hat m^*_0 &=& \gamma_2, \label{Four_m2}
\end{eqnarray}\label{Four_m1_m2}\end{subequations}
with the right-hand sides
\begin{subequations}
\begin{eqnarray}
\gamma_1 &=& \Four\left\{K*(h_1m_0)\right\}\\
\gamma_2 &=& \Four\left\{K*(h_2m_0)\right\}
\end{eqnarray}\end{subequations}
and the abbreviation
\begin{eqnarray}
M = (1-\mu)\sigma^2 \label{abbrev_M}.
\end{eqnarray}
If $\DDrift$ and $\DDiff$ can be expressed as polynomials in $x$, Eqs.~(\ref{Four_m1_m2}) can be used to extract the
corresponding polynomial coefficients. Let $\DDrift$ and $\DDiff$ be given by
\begin{subequations}
\begin{eqnarray}
\DDrift &=& \sum_{j=0}^{N_1} a_j x^j\\
\DDiff &=& \sum_{j=0}^{N_2} b_j x^j.
\end{eqnarray}\label{def_polycoeff}\end{subequations}
Choosing the first order approximation Eqs.~(\ref{expansion_moments_euler}) for
$h_1$ and $h_2$, the right-hand sides of Eqs.~(\ref{Four_m1_m2}) read
\begin{subequations}
\begin{eqnarray}
\gamma_1 &=& \tau\Four\left\{K*(\DDrift m_0)\right\}\cr
\cr
         &=& \tau\sum_{j=0}^{N_1} a_j\Four\left(K*(x^j m_0)\right)\\
\cr
\gamma_2 &=& \tau\Four\left\{K*(\DDiff m_0)\right\}.\cr
\cr
         &=& \tau\sum_{j=0}^{N_2} b_j\Four\left(K*(x^j m_0)\right)
\end{eqnarray}\end{subequations}
So for polynomial drift and diffusion functions the problem of expressing $\gamma_1$ and $\gamma_2$ reduces to
finding expressions for the transforms of $K*(x^j m_0)$.
It should be noted here that this would also be the case for higher order approximations $h_1$ and $h_2$.

Again because $K$ is a Gauss function with variance $\sigma^2$, the Fourier transform of $K*(x^j m_0)$ can be
expressed as linear combination of the transforms of $x^k(K*m_0)$ (see appendix~\ref{app:gauss}). Since
$K*m_0=m_0^*$ these transforms can be derived from the noisy data. Using the
shortcuts
\begin{eqnarray}
\Phi_j &=& \Four\left( x^j(K*m_0) \right) = \Four\left( x^jm_0^* \right)\\
F_j &=& \Four\left( K*(x^jm_0) \right)
\end{eqnarray}
the relation between $\Phi$ and $F$ is given by
\begin{eqnarray}
F_j &=& \sum_{k=0}^j{j\choose k}\varphi_{j-k} \Phi_k
\end{eqnarray}
with
\begin{eqnarray}
\varphi_j=i^j\sum_{k=0}^{j}|a_{jk}|\sigma^{j+k}\omega^k
\end{eqnarray}
and
\begin{eqnarray}
\begin{array}{lll}
a_{jk} &=& -a_{(j-1)(k-1)}+(k+1)a_{(j-1)(k+1)}\\
a_{00} &=& 1\\
a_{jk} &=& 0\qquad (j<0,k<0,j<k).
\end{array}
\end{eqnarray}
Eqs.~(\ref{Four_m1_m2}) now read
\begin{subequations}
\begin{eqnarray}
\hat m^*_1 - M i\omega\hat m^*_0  = \tau\sum_{j=0}^{N_1} a_j F_j\qquad&&\\
\cr
\hat m^*_2 - 2M(\hat m^*_0 +i\omega\hat m^*_1)-M^2\omega^2\hat m^*_0 \qquad&&\cr
\cr
=\tau\sum_{j=0}^{N_2} b_j F_j.\qquad&&
\end{eqnarray}\label{final}\end{subequations}
If the noise parameters have been extracted according to section~\ref{sec:extracting noise}, then $M=(1-\mu)\sigma^2$
and $F_j$ are known and the coefficients $a_j$ and $b_j$ can be extracted by a least square fit. It is also possible to use
Eqs.~(\ref{final}) to fit noise parameters and polynomial coefficients simultaneously by an iterated
least square fit. But this has turned out to be less accurate.

Adding an increment dependency to the polynomial coefficients however allows some of the approximation errors to be absorbed by the
additional parameters and has shown to increase accuracy.
\begin{subequations}
\begin{eqnarray}
a_j &\to& a_j +\tau a_j^1\\
b_j &\to&  b_j +\tau b_j^1
\end{eqnarray}\end{subequations}
With this extension the final equations for parameter fitting are given by
\begin{subequations}
\begin{eqnarray}
\hat m^*_1 - M i\omega\hat m^*_0  = \tau\sum_{j=0}^{N_1} (a_j+\tau a_j^1) F_j\qquad&&\\
\cr
\hat m^*_2 - 2M(\hat m^*_0 +i\omega\hat m^*_1)-M^2\omega^2\hat m^*_0 \qquad&&\cr
\cr
=\tau\sum_{j=0}^{N_2} (b_j+\tau b_j^1) F_j.\qquad&&
\end{eqnarray}\label{final_t}\end{subequations}
To be able to actually perform a fit, a finite number of values for $\omega$ and $\tau$ has to be chosen.
It is sufficient to restrict the choice of $\omega$ to positive values because the above equations are symmetric
(conjugate complex) in $\omega$. The following heuristic approach has shown to work quite well:
\begin{itemize}
\item Define an upper bound $\omega^*$ by
\begin{eqnarray}
\int_0^{\omega^*}|\hat m_0^*|^2d\omega &=& 0.99 \int_0^\infty|\hat m_0^*|^2d\omega.
\end{eqnarray}
\item Chose $N_\omega$ values $\omega_j$ equally distributed in $[0,\omega^*]$. The actual number of
values is of minor importance but it should be high enough to properly sample $\hat m_0^*$. 
\end{itemize}

\section{Application to numerical data}
\label{sec:examples}

In order to test the accuracy of the proposed method, synthetic stochastic signals have been generated by numerical
integration. As a first test-case an Ornstein-Uhlenbeck process has been chosen.
Its drift and diffusion functions have been defined as
\begin{subequations}
\begin{eqnarray}
\DDrift(x) &=& -x\\
\DDiff(x) &=& 2.
\end{eqnarray}\end{subequations}
The generated time series will be referred to as data set A.
It consists of $10^6$ data points at time increments $\Delta t=10^{-2}$. The integration has been performed
with a timestep size of $10^{-4}$ using the Euler scheme.
The process-timescale of $1.0$ corresponds to $\tau=100$ in units of increments of data points.

\begin{figure}[t]
\begin{center}
\includegraphics*[width=8.6cm,angle=0]{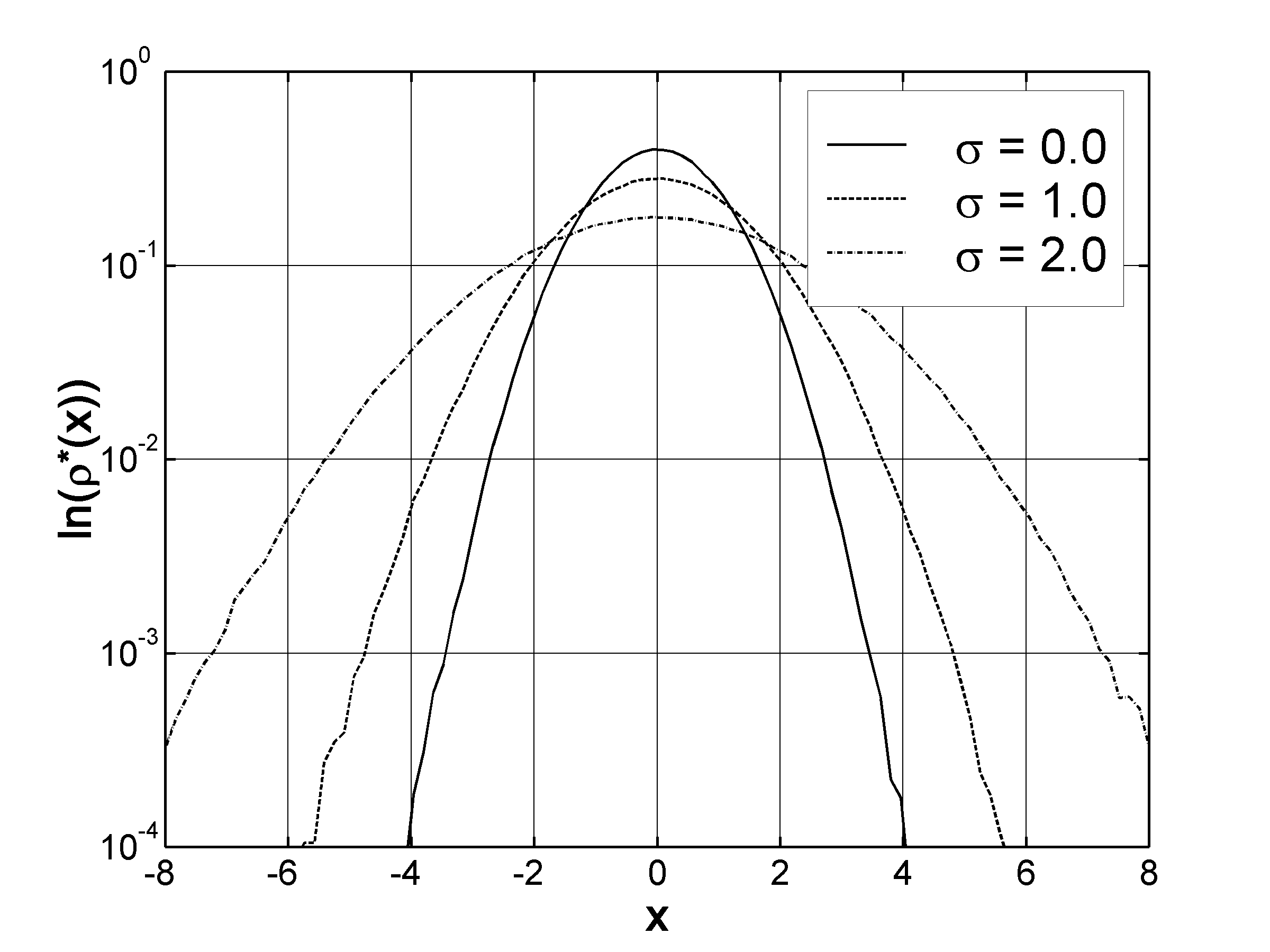}
\end{center}
\caption{\protect Probability density function for data set A. $\sigma$ is the
standard deviation of the superimposed noise. Drift and diffusion functions are given by $\DDrift=-x$ and $\DDiff=2$.
}
\label{fig1}
\end{figure}

This time series has been superimposed by additional measurement noise with standard deviation in the range
$\sigma=0.0\ldots 2.0$ and correlation time $T=0.0$ (delta-correlated). Some of the density histograms of the
resulting signals are shown in Fig. \ref{fig1}.

\begin{figure}[t]
\begin{center}
\includegraphics*[width=8.6cm,angle=0]{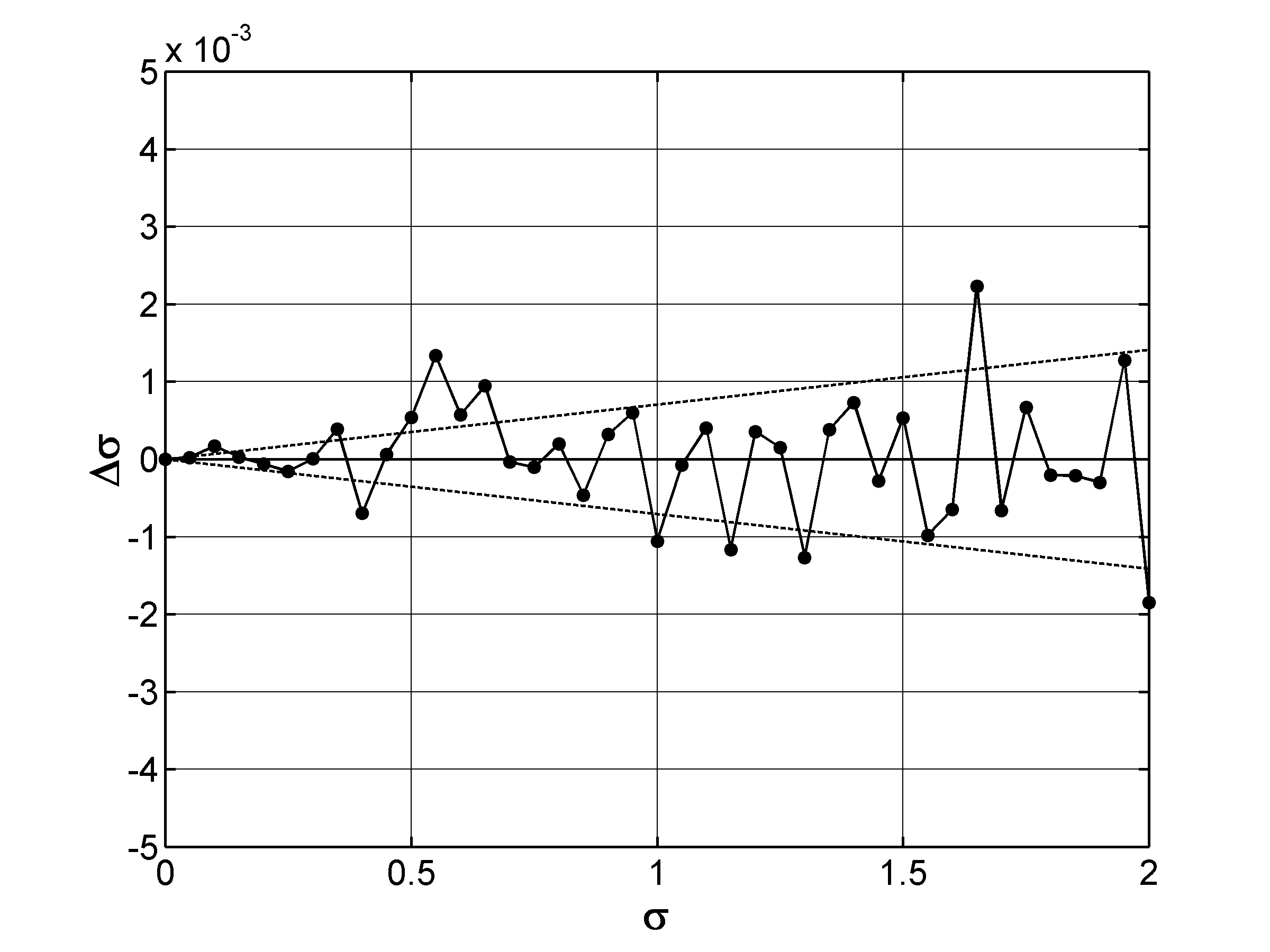}
\end{center}
\caption{\protect Absolute error $\Delta\sigma=\tilde\sigma-\sigma$ in the estimation of the measurement noise
of data set A. The dashed lines indicate the expected lower bound of the
standard deviation of $\Delta\sigma$.
}
\label{fig3}
\end{figure}

The noise parameters have been extracted according to section \ref{sec:extracting noise}. The maximum increment has
been $\tau=60$. The absolute error $\Delta\sigma$ in the estimation of the
strength of the measurement noise is presented in Fig. \ref{fig3} as a function of the true measurement noise $\sigma$.
The results are accurate within the achieveable precision of the finite time series. This means that the measurement noise alone,
without any stochastic process, cannot be estimated much more accurate: For a sample of $N$ data points of
Gaussian, delta-correlated noise the estimation $S$ of the true standard deviation, $\sigma$, obeys a variance of approximately
$\sigma^2\frac{1}{2N}$ (see appendix~\ref{app:err_sigma}). A lower bound for the standard deviation of $\Delta\sigma$ is therefore given by
$\sigma\frac{1}{\sqrt{2N}}$. These limits are indicated in Fig. \ref{fig3} as dashed lines.

\begin{figure}[tb]
\begin{center}
\includegraphics*[width=8.6cm,angle=0]{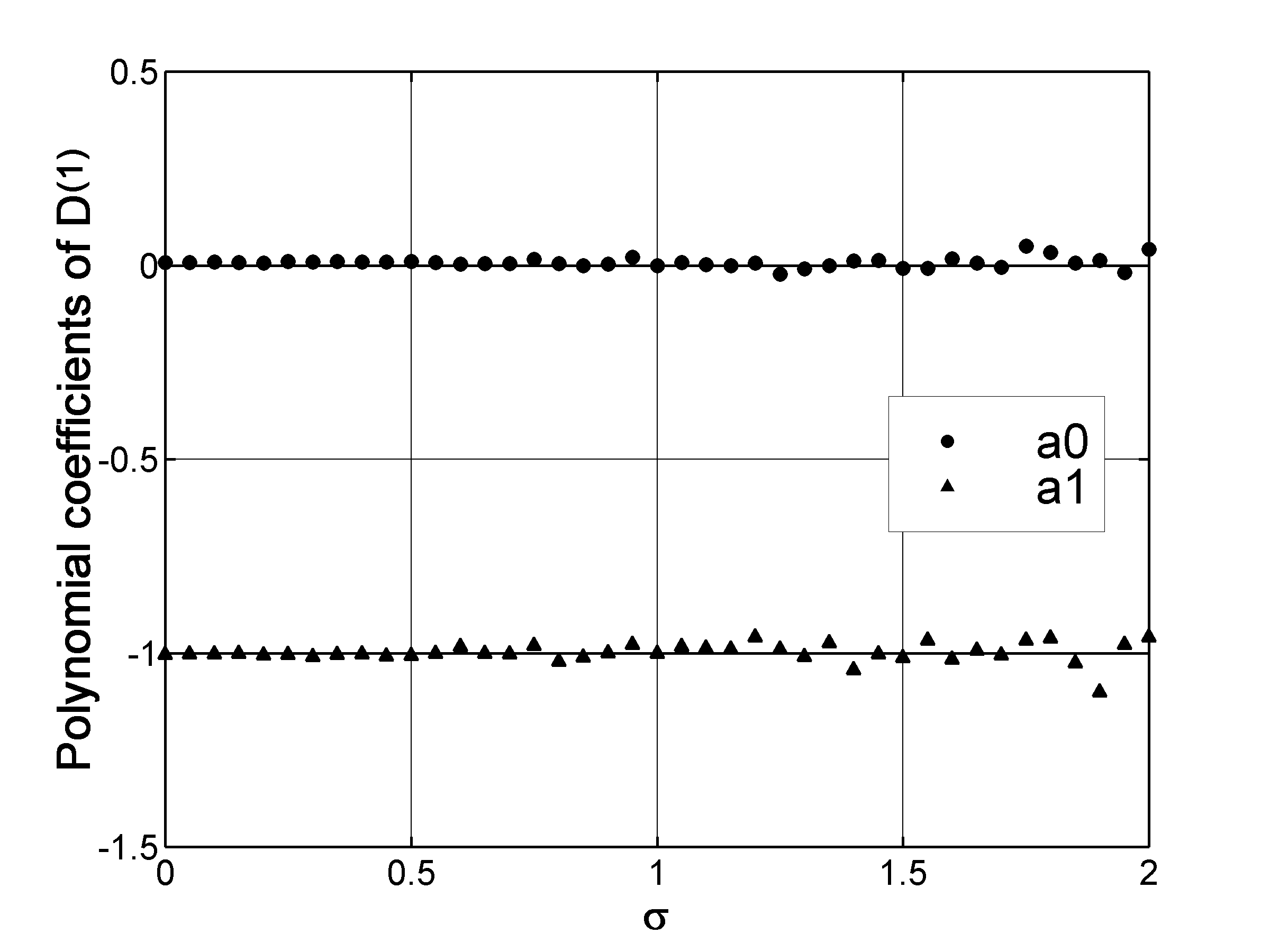}
\end{center}
\caption{\protect Estimated polynomial coefficients of $\DDrift$ for data set A
with $a_0$ and $a_1$ according to Eq.~(\ref{def_polycoeff}).
}
\label{fig4}
\end{figure}

\begin{figure}[tb]
\begin{center}
\includegraphics*[width=8.6cm,angle=0]{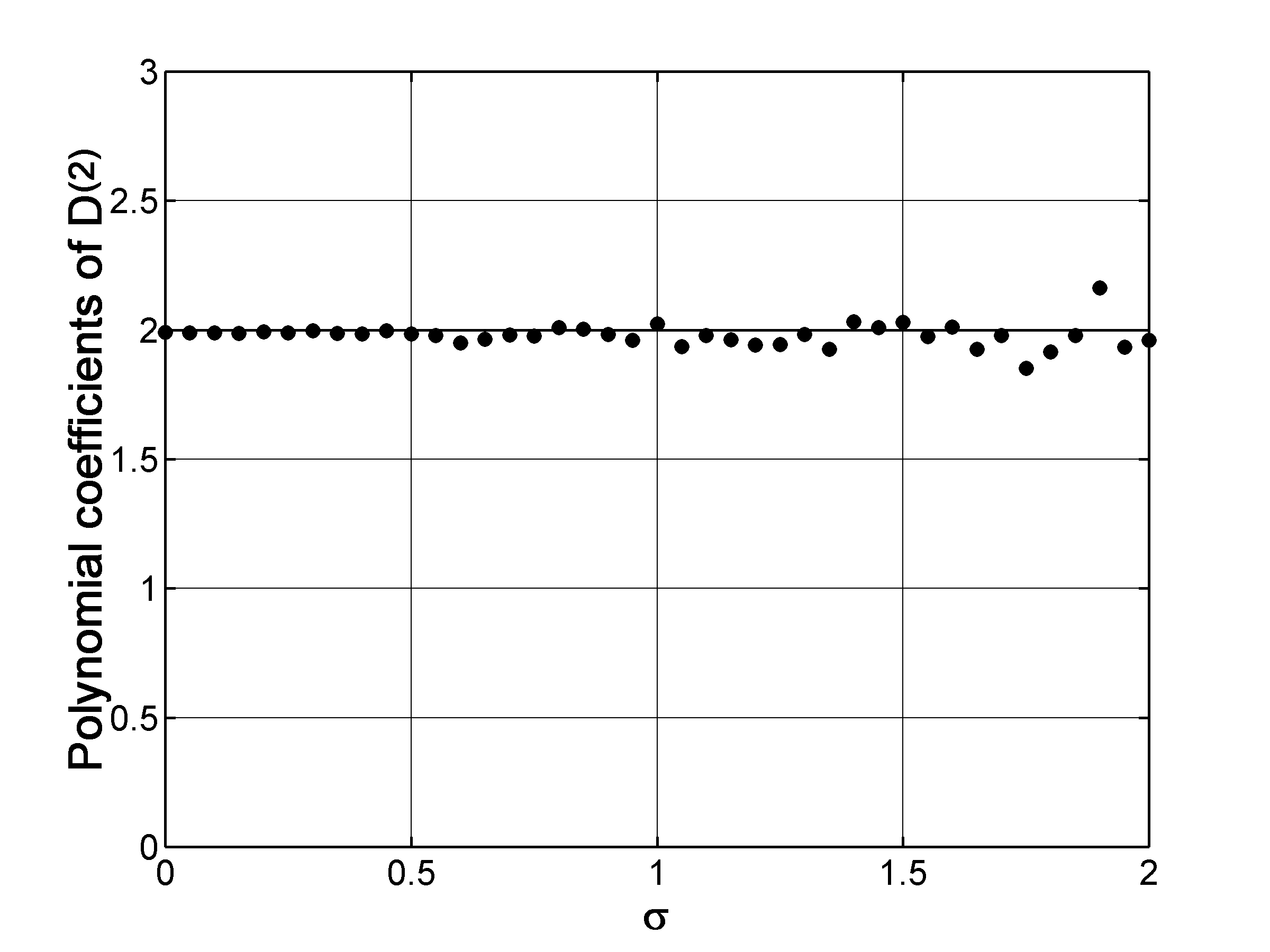}
\end{center}
\caption{\protect Estimated polynomial coefficient $b_0$ of $\DDiff$ for data set A
with $b_0$ according to Eq.~(\ref{def_polycoeff}).
}
\label{fig5}
\end{figure}

Next the drift and diffusion functions have been fitted choosing the polynomial ansatz $\DDrift=a_0+a_1x$ and
$\DDiff=b_0$. As maximum increment a value of $\tau=25$ (again in units of increments of data points) has been chosen.
The results are shown in Figs. \ref{fig4} and \ref{fig5} and are in good aggreement with the true values ($a_0=0$, $a_1=-1$,
$b_0=2$).

After this basic test-case a more general case has been chosen.
It includes multiplicative noise, heavy tails and a density distribution which is not symmetric.
Also the superposition of noise with finite
correlation time has been investigated in this setup. Drift and diffusion functions have been defined as
\begin{subequations}
\begin{eqnarray}
\DDrift(x) &=& 1-x\\
\DDiff(x) &=& 2-2x+2x^2.
\end{eqnarray}\end{subequations}
The generated time series will be referred to as data set B.
Again it consists of $10^6$ data points at time increments $\Delta t=10^{-2}$. The integration has been performed
with a timestep size of $2.0\cdot10^{-5}$ using the Euler scheme. The standard deviation of the data points
is approximately $1.83$
and the timescale of the process has been estimated via the autocorrelation function to be about $\tau=100$ in units
of increments of data points.

\begin{figure}[t]
\begin{center}
\includegraphics*[width=8.6cm,angle=0]{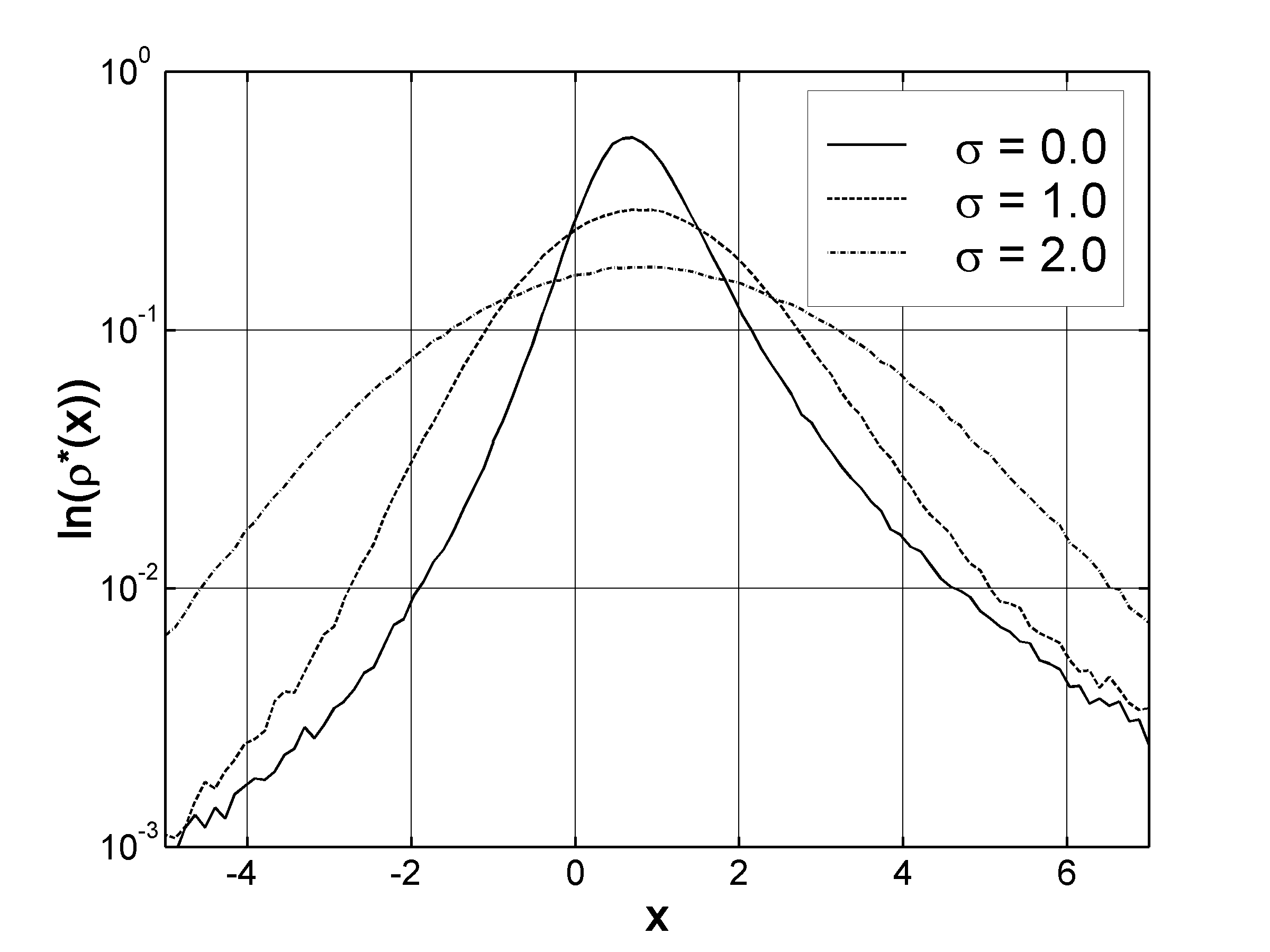}
\end{center}
\caption{\protect Probability density function of data set B. $\sigma$ is the
standard deviation of the superimposed noise. Drift and diffusion functions are
given by $\DDrift=1-x$ and $\DDiff=2-2x+2x^2$.
}
\label{fig6}
\end{figure}

In a first step again delta-correlated measurement noise with standard deviation in the range
$\sigma=0.0\ldots 2.0$ has been superimposed to the process. Some of the density histograms of the
resulting signals are shown in Fig. \ref{fig6}.

\begin{figure}[t]
\begin{center}
\includegraphics*[width=8.6cm,angle=0]{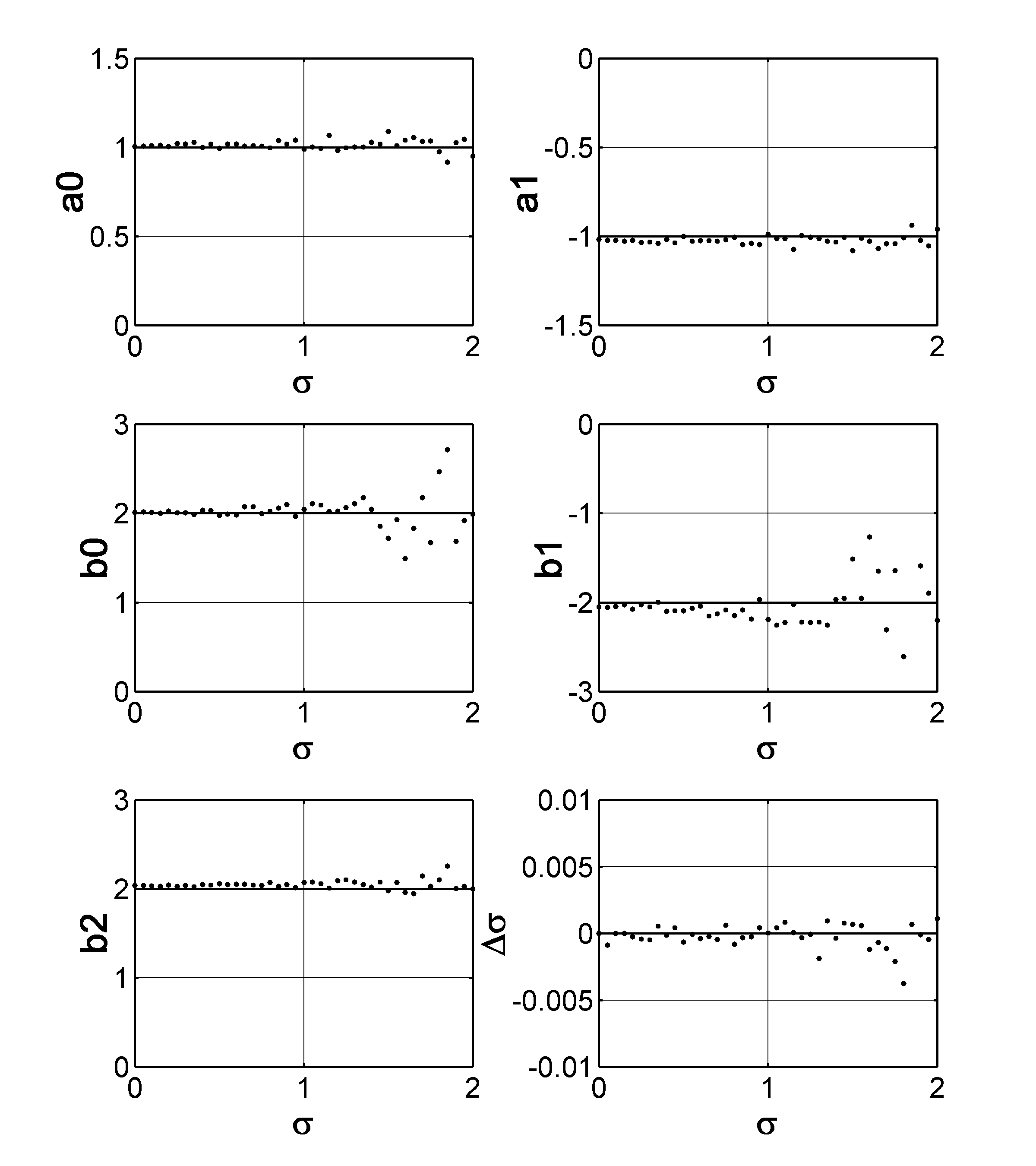}
\end{center}
\caption{\protect Results for data set B. Estimations of the polynomial coefficients $a_j$ and $b_j$
and absolute error $\Delta\sigma$ in the estimation of the measurement noise.
The superimposed noise is delta-correlated.
}
\label{fig7}
\end{figure}
As maximum increment for noise-fitting $\tau=60$ and for coefficient-fitting $\tau=25$ has been chosen.
The polynomial ansatz for drift and diffusion functions is given by $\DDrift=a_0+a_1x$ and
$\DDiff=b_0+b_1x+b_2x^2$. The results are summarized in Fig. \ref{fig7}.

\begin{figure}[t]
\begin{center}
\includegraphics*[width=8.6cm,angle=0]{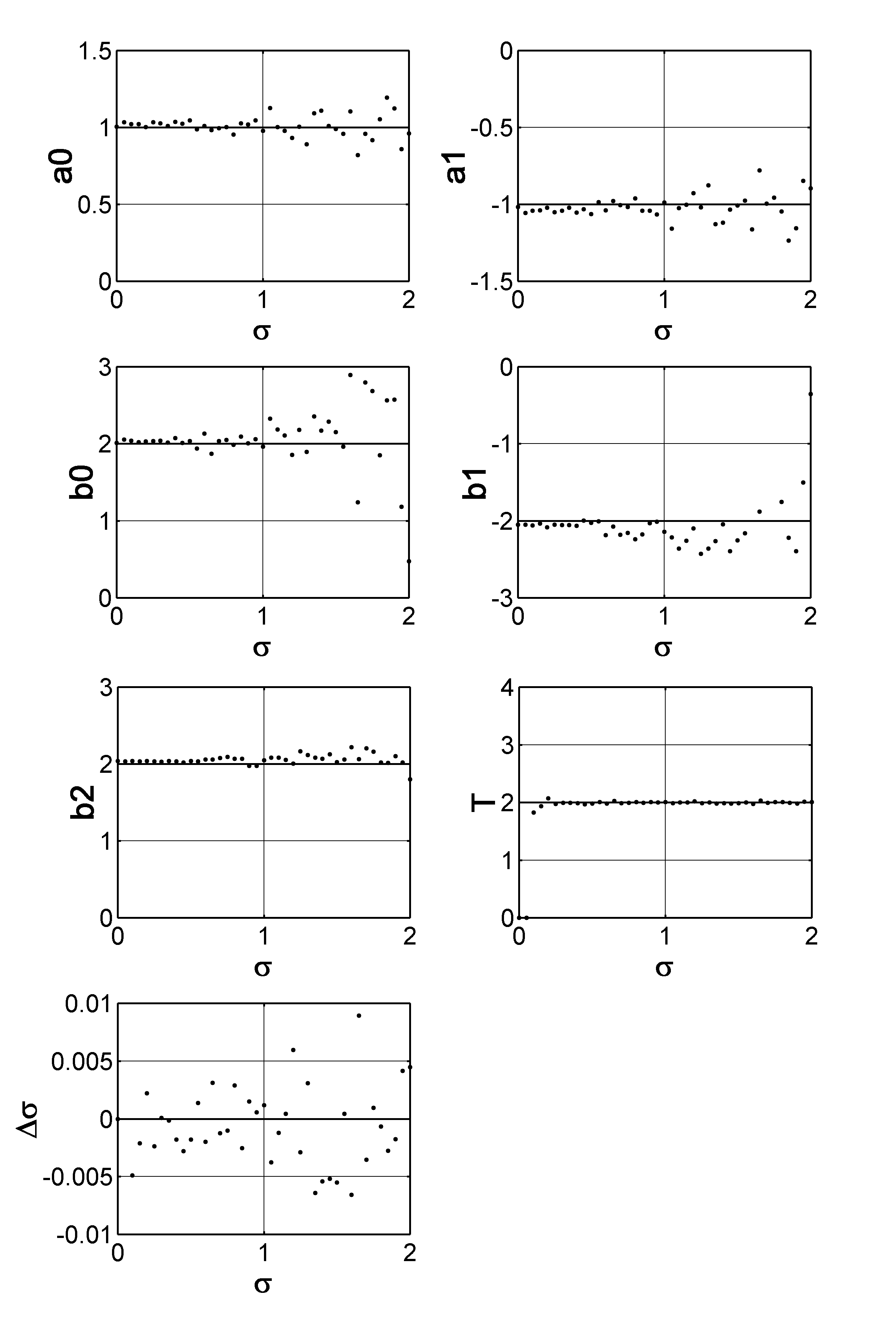}
\end{center}
\caption{\protect Results for data set B. Estimations of the polynomial coefficients $a_j$ and $b_j$,
correlation time $T$ and absolute error $\Delta\sigma$ in the
estimation of the measurement noise.
The superimposed noise has a correlation time of $T=2.0$.
}
\label{fig8}
\end{figure}

\begin{figure}[t]
\begin{center}
\includegraphics*[width=8.6cm,angle=0]{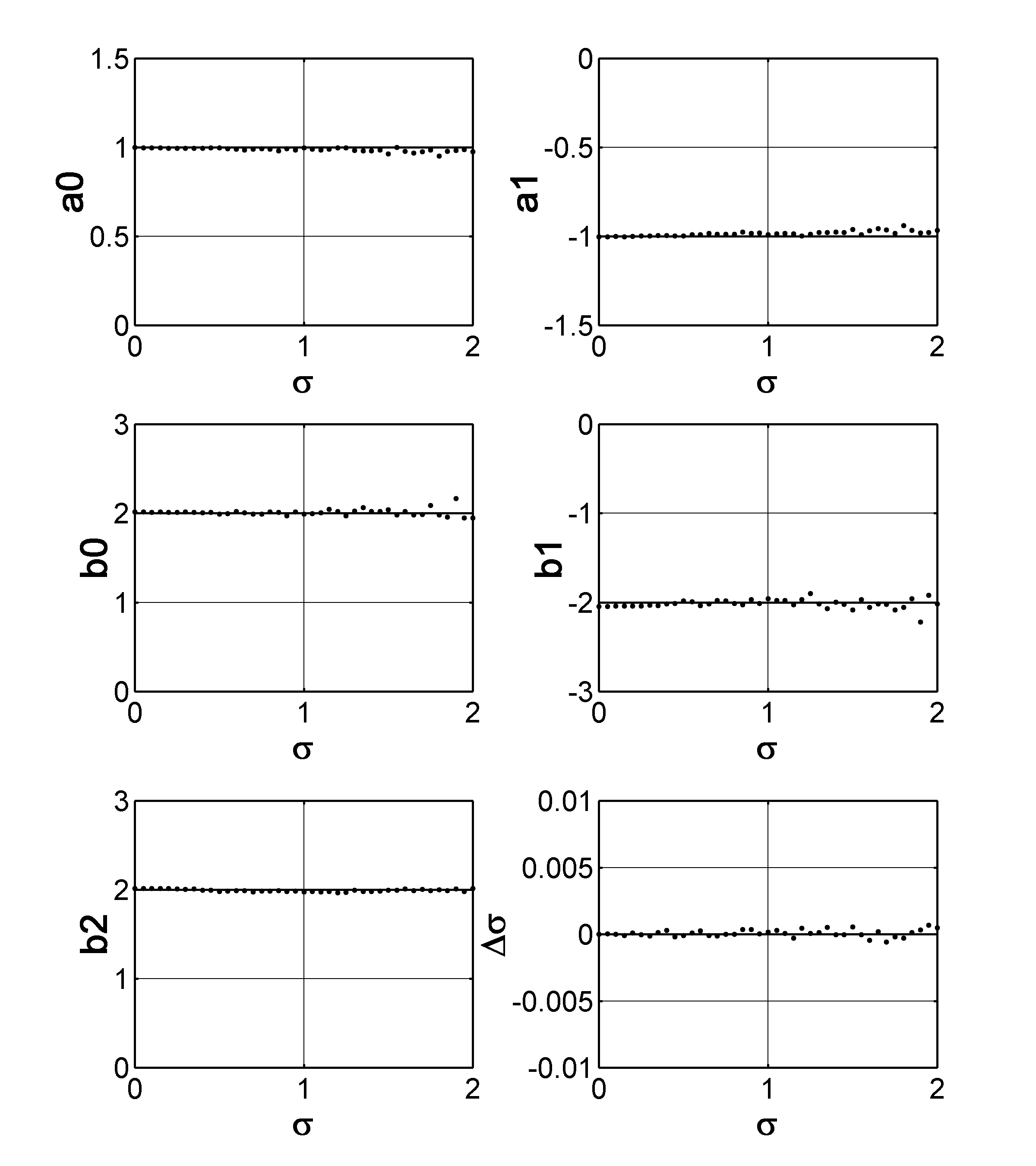}
\end{center}
\caption{\protect Results for data set B1. Estimations of the polynomial coefficients $a_j$ and $b_j$
and absolute error $\Delta\sigma$ in the estimation of the measurement noise.
The superimposed noise is delta-correlated.
}
\label{fig9}
\end{figure}

\begin{figure}[htb]
\begin{center}
\includegraphics*[width=8.6cm,angle=0]{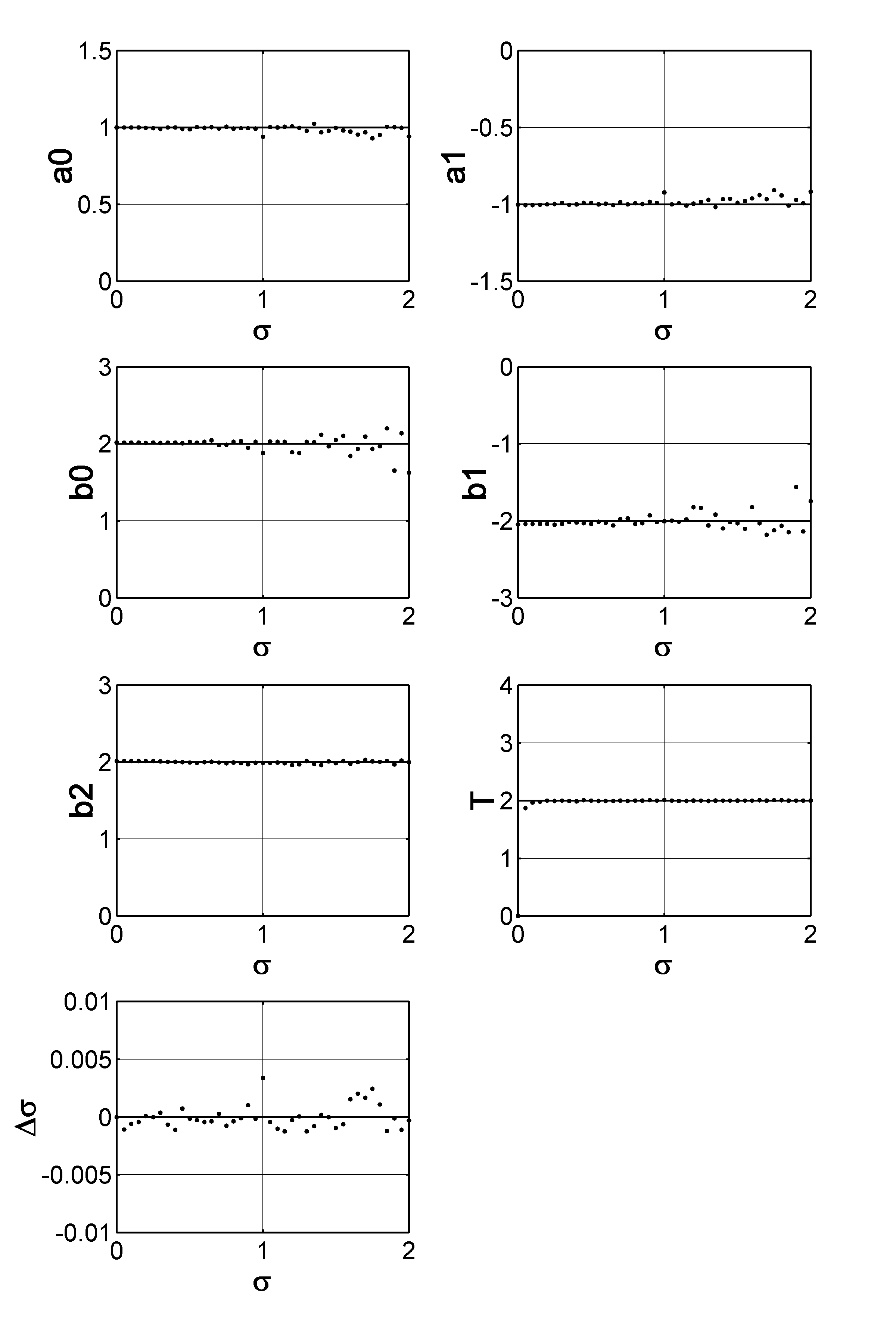}
\end{center}
\caption{\protect Results for data set B1. Estimations of the polynomial coefficients $a_j$ and $b_j$,
correlation time $T$ and absolute error $\Delta\sigma$ in the
estimation of the measurement noise.
The superimposed noise has a correlation time of $T=2.0$.
}
\label{fig10}
\end{figure}

In a second step measurement noise with a correlation time $T=2.0$ (in units of data point increments) has been superimposed.
The results are given in Fig. \ref{fig8} and obey higher fluctuations as in the delta-correlated case.
The correlation time of the measurement noise is estimated nicely except for small amplitudes of $\sigma$. To check
the assumption that this is caused by the finite-sample fluctuations of $z(\tau)$ a larger data set, B1, has been generated.
It consists of $10^7$ data points and its analysis should show less variability in all estimated quantities. This is indeed
the case. The results are shown in Figs. \ref{fig9} and \ref{fig10}.


\section{Conclusions}
\label{sec:conclusions}

A new procedure has been described to analyze stochastic time series that are superimposed by strong
measurement noise. The algorithm is able to cope with exponentially correlated noise
and accurately extracts strength and correlation time of the measurement noise as well as the parameters defining the drift
and diffusion functions of the underlying stochastic process. This has been shown by the analysis of synthetically
generated time series. The chosen stochastic processes include the cases of multiplicative noise and heavy tailed density
distribution.

The computational costs of the algorithm are quite low. It takes less than a minute to analyze a data set of size
$10^7$ data points on a usual desktop PC.

A first order Taylor-It\^o expansion for the moments of the finite time differences of the process variable is used
in the current implementation. It is straightforward to extend the algorithm to also take higher order terms into
account. This should extend the range of time increments that can be used for the analysis and thus should increase
the accuracy of the results. This has to be done in the future.

Another future task is the extension to higher dimensional processes. Up to now it is an open question if a
general
multidimensional Ornstein-Uhlenbeck process as a model for the measurement noise can be used
or if some restrictions need to be imposed on it.

\section{Acknowledgments}
\label{sec:ack}

The author especially wants to thank Joachim Peinke, Rudolf Friedrich, David Kleinhans, Pedro G. Lind
and Maria Haase for many useful discussions.

\appendix
\section{Gauss functions}
\label{app:gauss}
\subsection{Definitions and basic properties}
\label{app:gauss:basic}

Let Fourier transform and convolution be defined as below. For notational simplicity the `hat' syntax
will be used to denote the transform of single functions. For more complex expressions the functional form
${\bf{F}}(\ldots)$ will usually be the better choice.
\begin{eqnarray}
\Four(f(x)) &=& \hat{f}(\omega) = \int_{-\infty}^{+\infty}\hbox{e}^{-i\omega x}f(x)dx\\
f(x)*g(x) &=& \int_{-\infty}^{+\infty}f(x-x^\prime)g(x^\prime) dx
\end{eqnarray}
Above definitions imply the following properties:
\begin{subequations}
\begin{eqnarray}
\Four(\p_x^jf(x)) &=&  (i\omega)^j\hat{f} (\omega) \label{basic:prop1a}\\
&&\cr
\Four(x^jf(x)) &=& (i\p_\omega)^j\hat{f}(\omega) \label{basic:prop1b}\\
&&\cr
\Four(f(\frac{x}{a})) &=& |a|\hat{f}(a\omega) \label{basic:prop1c}\\
&&\cr
\Four(f(x)*g(x)) &=& \hat{f}(\omega)\hat{g}(\omega) \label{basic:prop1d}\\
&&\cr
\p_x^j(f(x)*g(x)) &=& (\p_x^jf(x))*g(x)\cr\cr
                  &=& f(x)*(\p_x^jg(x)) \label{basic:prop1e}
\end{eqnarray}\label{basic:prop1}
\end{subequations}
%
\subsection{Derivatives and polynomial products of Gauss functions}
\label{app:gauss:deriv}

Let the Gauss function $G$ be defined as below.
\begin{subequations}
\begin{eqnarray}
G(x) &=& \frac{1}{\sqrt{2\pi}} \hbox{e}^{-\frac{1}{2}x^2} \label{basic:gauss1a}\\
&&\cr
\hat{G}(\omega) &=& \hbox{e}^{-\frac{1}{2}\omega^2}= \sqrt{2\pi}G(\omega) \label{basic:gauss1b}
\end{eqnarray}\label{basic:gauss1}\end{subequations}
All derivatives  $\p_x^jG(x)$ have the form $P_j(x)G(x)$, where $P_j$ is a polynomial of order $j$ in $x$.
\begin{eqnarray}
\p_xG(x)   &=& \left\{-x\right\}G(x)\cr
\cr
\p_x^2G(x) &=& \left\{-1+x^2\right\}G(x)\cr
           &\vdots& \cr
\p_x^jG(x) &=& \left\{\sum_{k=0}^ja_{jk}x^k\right\}G(x)\label{deriv:def1}
\end{eqnarray}
\begin{eqnarray*}
a_{jk} &=&\left(
\begin{array}{rrrrrr}
 1&  &  &  &\phantom{-1}& \\
 0&-1&  &  &  & \\
-1& 0& 1&  &  & \\
 0& 3& 0&-1&  & \\
 3& 0&-6& 0& 1& \\
 \vdots&  &  &  &  &\ddots \\
\end{array}
\right)
\end{eqnarray*}
The values of the polynomial coefficients can be determined by a recursion formula:
\begin{eqnarray}
\begin{array}{lll}
a_{jk} &=& -a_{(j-1)(k-1)}+(k+1)a_{(j-1)(k+1)}\\
a_{00} &=& 1\\
a_{jk} &=& 0\qquad (j<0,k<0,j<k)
\end{array}\label{deriv:recursion}
\end{eqnarray}
It can also be shown, that the coefficients are zero, if $(j+k)$ or $(j-k)$ is odd. Otherwise, within the restrictions
of formula~(\ref{deriv:recursion}),  $a_{jk}$ is non-zero and its sign is given by $i^{j+k}$.

It will be useful later, to express the derivatives of $\frac{1}{G}$ in a similar way. The proceeding is the same as
for $G$ and leads to the recursion formula~(\ref{deriv:recursion2}).
\begin{eqnarray}
\p_x^j\frac{1}{G(x)} &=& \left\{\sum_{k=0}^j\tilde{a}_{jk}x^k\right\}\frac{1}{G(x)}\label{deriv:def2}
\end{eqnarray}
\begin{eqnarray}
\begin{array}{lll}
\tilde{a}_{jk} &=& \tilde{a}_{(j-1)(k-1)}+(k+1)\tilde{a}_{(j-1)(k+1)}\\
\tilde{a}_{00} &=& 1\\
\tilde{a}_{jk} &=& 0\qquad (j<0,k<0,j<k) \phantom{\int_\int}
\end{array}\label{deriv:recursion2}
\end{eqnarray}
Having a closer look at the coefficients, it turns out, that  $\tilde{a}_{jk}$ and $a_{jk}$ only differ in sign.
\begin{eqnarray}
\tilde{a}_{jk}=|a_{jk}|\label{deriv:def2a}
\end{eqnarray}
To get explicit expressions  for $x^jG(x)$ in terms of derivatives of $G$, a Fourier transformation is applied
to Eq.~(\ref{deriv:def1}).
\begin{eqnarray}
(i\omega)^j\hat G(\omega) &=& \left\{\sum_{k=0}^ja_{jk}(i\p_\omega)^k\right\}\hat G(\omega).\label{deriv:def3}
\end{eqnarray}
Using Eq.~(\ref{basic:gauss1b}) and substituting $\omega$ by $x$, finally yields:
\begin{eqnarray}
x^j G(x) &=& \left\{\sum_{k=0}^j i^{k-j} a_{jk}\p_x^k\right\} G(x).\label{deriv:def4}
\end{eqnarray}
%
\subsection{Scaled Gauss functions}
\label{app:gauss:scaled}

This can also be expressed for scaled Gauss functions having standard deviation $\sigma$. The
Fourier transform of such a function is given by
\begin{eqnarray}
\Four{(G(\xosig})) &=& \sigma\hat G(\sigom).   \label{scaled:G}
\end{eqnarray}
The relations between derivatives and polynomial products then read:
\begin{eqnarray}
\p_x^jG(\xosig) &=& \left\{\sum_{k=0}^ja_{jk}\sigma^{-j-k}x^k\right\}G(\xosig)\label{app:gauss:scaled:dG}\\
\p_x^j\frac{1}{G(\xosig)} &=& \left\{\sum_{k=0}^j|a_{jk}|\sigma^{-j-k}x^k\right\}\frac{1}{G(\xosig)}\label{app:gauss:scaled:dG_inv}\\
x^j G(\xosig) &=& \left\{\sum_{k=0}^j i^{k-j} a_{jk}\sigma^{j+k}\p_x^k\right\} G(\xosig)\label{app:gauss:scaled:xG}
\end{eqnarray}
%
\begin{widetext}
\subsection{Convolutions}
\label{app:gauss:conv}

The above properties can be used to express convolutions of the form $(x^jG)*f$ and $G*(x^jf)$ in terms of the `raw'
convolution $G*f$ and its polynomial products $x^j(G*f)$. Using Eqs.~(\ref{basic:prop1e}) and (\ref{app:gauss:scaled:xG})
immediately leads to Eq.~(\ref{conv:xGf}).
\begin{eqnarray}
(x^j G(\xosig))*f(x) &=& \left(\left\{\sum_{k=0}^j i^{k-j} a_{jk}\sigma^{j+k}\p_x^k\right\}
                         G(\xosig)\right)*f(x)\cr
                     &=& \left\{\sum_{k=0}^j i^{k-j} a_{jk}\sigma^{j+k}\p_x^k\right\}
                         (G(\xosig)*f(x) ) \quad\label{conv:xGf}
\end{eqnarray}
To get an expression for $G*(x^jf)$ it is useful to look at the term $G\cdot\p^jf$ first.
\begin{eqnarray*}
G(\xosig)\p_x^jf(x) &=& G(\xosig)\p_x^j\left\{\frac{1}{G(\xosig)}G(\xosig)f(x)\right\}\cr
   &=& G(\xosig)\sum_{k=0}^j {j\choose k}\left\{\p_x^{j-k}\frac{1}{G(\xosig)}\right\} \p_x^k\left\{G(\xosig)f(x)\right\}\cr
   &=& G(\xosig)\sum_{k=0}^j {j\choose k}\left\{\sum_{l=0}^{j-k}|a_{(j-k)l}|\sigma^{-j+k-l}x^l\right\}\frac{1}{G(\xosig)} \p_x^k\left\{G(\xosig)f(x)\right\}\cr
   &=& \sum_{k=0}^j {j\choose k}\sum_{l=0}^{j-k}|a_{(j-k)l}|\sigma^{-j+k-l}x^l \p_x^k\left\{G(\xosig)f(x)\right\}\cr
\end{eqnarray*}
This allows the transform of $G*(x^jf)$ to be written as follows:
\begin{eqnarray}
\Four\left( G(\xosig)*(x^jf(x)) \right) &=& \sigma\hat G(\sigom)(i\p_\omega)^j\hat f(\omega)
                                         = i^j\sigma\sqrt{2\pi}\cdot G(\sigom)\p_\omega^j\hat f(\omega)\cr
   &=& i^j\sigma\sqrt{2\pi}\sum_{k=0}^j{j\choose k}\sum_{l=0}^{j-k}|a_{(j-k)l}|\sigma^{j-k+l}\omega^l
     \p_\omega^k\left\{ G(\sigom)\hat f(\omega) \right\}\cr
   &=& \sum_{k=0}^j{j\choose k}\sum_{l=0}^{j-k}i^{j-k}|a_{(j-k)l}|\sigma^{j-k+l}\omega^l
     (i\p_\omega)^k\left\{ \sigma\hat G(\sigom)\hat f(\omega) \right\}\cr
\cr
  &=&\sum_{k=0}^j{j\choose k}\sum_{l=0}^{j-k}i^{j-k}|a_{(j-k)l}|\sigma^{j-k+l}
        \omega^l\Four\left( x^k(G(\xosig)*f(x)) \right)\ \label{conv:FGxf}
\end{eqnarray}
Omitting arguments and introducing the auxiliary quantity $\varphi_j$ this can be expressed as
\begin{eqnarray}
\Four\left( G*(x^jf) \right) &=& \sum_{k=0}^j{j\choose k}\varphi_{j-k}\Four\left( x^k(G*f) \right)
                            ,\qquad \varphi_j=i^j\sum_{k=0}^{j}|a_{jk}|\sigma^{j+k}\omega^k.
\end{eqnarray}
%

\section{Conditioned moments $m^*_j$}
\label{app:cond_mom_1}

After inserting Eqs.~(\ref{pdf_noise}) and (\ref{pdf_noisy}) and interchanging the order
of integration Eq.~(\ref{m_j_eq_1}) reads
\begin{eqnarray}
m_j^*(x_1,\tau) &=& \int_{-\infty}^\infty K(x_1-x^\pr_1)
                    \int_{-\infty}^\infty \rho(x^\pr_1,x^\pr_2,\tau) I_j dx^\pr_2 dx^\pr_1\label{app:m_j_raw},
\end{eqnarray}
where
\begin{eqnarray}
I_j &=& \int_{-\infty}^\infty (x_2-x_1)^j \frac{1}{\sqrt{2\pi s^2}}
       e^{ -\frac{1}{2} \left\{\frac{x_2-\mu(\tau)(x_1-x^\pr_1)-x^\pr_2}{s}\right\}^2  }dx_2. \label{app:I_j}
\end{eqnarray}
Substituting
\begin{eqnarray}
z &=& \frac{x_2-\mu(\tau)(x_1-x^\pr_1)-x^\pr_2}{s}
\end{eqnarray}
gives $I_j = \int_{-\infty}^\infty (sz +(x^\pr_2-x^\pr_1)-(1-\mu(\tau))(x_1-x^\pr_1))^j
\frac{1}{\sqrt{2\pi}} e^{ -\frac{1}{2} z^2  }dz$ and therefore
\begin{subequations}
\begin{eqnarray}
I_0 &=& 1\\
\cr
I_1 &=& (x^\pr_2-x^\pr_1)-(1-\mu(\tau))(x_1-x^\pr_1)\\
\cr
I_2 &=& s^2+(x^\pr_2-x^\pr_1)^2 -2(1-\mu(\tau))(x_1-x^\pr_1)(x^\pr_2-x^\pr_1)
                         +(1-\mu(\tau))^2(x_1-x^\pr_1)^2
\end{eqnarray}
\end{subequations}
Now $I^\pr_j = \int_{-\infty}^\infty \rho(x^\pr_1,x^\pr_2,\tau) I_j dx^\pr_2$ can be evaluated using the
definition of $m_j$ (Eq.~(\ref{m_j_no_noise})).
\begin{subequations}
\begin{eqnarray}
I^\pr_0 &=& m_0(x^\pr_1)\\
\cr
I^\pr_1 &=& m_1(x^\pr_1,\tau)-(1-\mu(\tau))(x_1-x^\pr_1)m_0(x^\pr_1)\\
\cr
I^\pr_2 &=& m_2(x^\pr_1,\tau) -2(1-\mu(\tau))(x_1-x^\pr_1)m_1(x^\pr_1,\tau)
                         +\left\{(1-\mu(\tau))^2(x_1-x^\pr_1)^2+s^2\right\}m_0(x^\pr_1)
\end{eqnarray}
\end{subequations}
Using $\int_{-\infty}^\infty(x-x^\pr)^kf(x-x^\pr)g(x^\pr)dx^\pr = (x^kf(x))*g(x)$ the noisy conditioned moments read
\begin{subequations}
\begin{eqnarray}
m^*_0(x_1)      &=& K(x_1)*m_0(x_1)\\
\cr
m^*_1(x_1,\tau) &=& K(x_1)*m_1(x_1,\tau)   -(1-\mu(\tau))(x_1K(x_1))*m_0(x_1)\\
\cr
m^*_2(x_1,\tau) &=& K(x_1)*m_2(x_1,\tau)    -2(1-\mu(\tau))(x_1K(x_1))*m_1(x_1,\tau)\cr
\cr
                          &&\quad  +(1-\mu(\tau))^2(x^2_1K(x_1))*m_0(x_1)+s^2K(x_1)*m_0(x_1).
\end{eqnarray}
\end{subequations}

\end{widetext}
\section{Error in the estimation of $\sigma$}
\label{app:err_sigma}

Let $\xi$ be a Gaussian random variable with mean zero and standard deviation $\sigma$. Further let $\xi_1,\ldots\xi_N$
be a sample of $N$ independent realizations of $\xi$. Then the expectation values of $\xi_i$, $\xi_i\xi_j$ and
$\xi_i^2\xi_j^2$ are given by
\begin{eqnarray}
<\xi_i> &=& 0 \\
<\xi_i\xi_j> &=& \sigma^2\delta_{ij} \label{xi2}\\
<\xi_i^2\xi_j^2> &=& \sigma^4+2\sigma^4\delta_{ij} .\label{xi4}
\end{eqnarray}
The variance of $\xi$ can be estimated from the sample as
\begin{eqnarray}
V &=& \frac{1}{N}\sum_{i=1}^N \xi_i^2
\end{eqnarray}
with an expectation value of $\sigma^2$ due to Eq.~(\ref{xi2}).
\begin{eqnarray}
<V> &=& \sigma^2
\end{eqnarray}
Therefore $V$ can be written as
\begin{eqnarray}
V &=& \sigma^2+\Delta V\qquad\hbox{with}\qquad <\Delta V> \;=\;0.
\end{eqnarray}
The variance of $\Delta V$ can be evaluated using Eq.~(\ref{xi4}).
\begin{eqnarray}
<(\Delta V)^2> &=& <(\frac{1}{N}\sum_{i=1}^N \xi_i^2-\sigma^2)^2> \cr
               &=& \frac{1}{N^2}\sum_{i=1}^N\sum_{j=1}^N <\xi_i^2\xi_j^2>-\sigma^4 \cr
               &=& \frac{1}{N^2}(N^2\sigma^4+2N\sigma^4)-\sigma^4 \cr
               &=& \frac{2}{N}\sigma^4
\end{eqnarray}
Up to now only an estimation of the {\em variance} of $\xi$ has been made. The estimation $S$ of its standard deviation
is given by
\begin{eqnarray}
S &=& \sqrt{V} \;=\; \sigma\sqrt{1+\frac{\Delta V}{\sigma^2}}
\end{eqnarray}
which, for large $N$, can be approximated as
\begin{eqnarray}
S &\approx& \sigma\cdot(1+\frac{1}{2}\frac{\Delta V}{\sigma^2}) \;=\; \sigma+\frac{\Delta V}{2\sigma} .
\end{eqnarray}
The standard deviation of the error in $S$ is therefore given by
\begin{eqnarray}
\Delta\sigma &=& \sqrt{<\left(\frac{\Delta V}{2\sigma}\right)^2>} \;=\; \frac{\sigma}{\sqrt{2N}}
\end{eqnarray}
%


%
\end{document}